\documentclass[%
 reprint,
 amsmath,amssymb,
 aps,
pra,
]{revtex4-2}

\usepackage{graphicx}
\usepackage{dcolumn}
\usepackage{bm}

\usepackage{soul}
\usepackage {caption}
\usepackage{times}
\usepackage{subcaption}
\usepackage{ragged2e}
\usepackage{float}
\usepackage{mathtools}
\usepackage{comment}
\newcommand{\thickhat}[1]{\mathbf{\hat{\text{$#1$}}}}
\usepackage{siunitx}
\DeclareMathOperator{\arccosh}{arcCosh}
\begin{document}

\preprint{APS/123-QED}

\title{Non-Classicality  and Non-adiabaticity   in   a Single Trapped Ion}

\author{C. F. P. Avalos}
\author{M. C. de Oliveira}%
 \email{Corresponding Author: marcos@ifi.unicamp.br}
\affiliation{%
 Gleb Wataghin Institute of Physics, University of Campinas, 13083-859, Campinas, São Paulo, Brazil.
}%




\date{\today}

\begin{abstract}
 Trapped ion systems present non-classical characteristics such as squeezed states that show a quantum advantage in quantum sensing, quantum information processing and quantum thermodynamics. We analyze the non-classical characteristics of a system described by a single ion trapped by a periodic potential field. Within the regime of non-adiabatic manipulation of the potential field, the dynamics of motion of the center of mass of the ion can be described by a dimensionless parameter called the non-adiabatic parameter $Q^{*}$. This parameter allows us to distinguish the classical and non-classical characteristics of the system. Using the equations of motion of observables in the Heisenberg picture, we propose an analysis of the unitary time evolution operator and discuss the squeezing behavior in the   state of motion of the ion. The results shown can serve as a basis to discuss the presence of squeezing as a resource in quantum thermodynamics in the non-adiabatic regime in actual achievable experimental limitations.

\end{abstract}

\maketitle


\section{\label{sec:level1} Introduction}
Trapped ions is a quantum system \cite{Paul1990traps, SeidelinSurface2006} that is present in several areas of physics such as: quantum computing \cite{CiracQuantumC1995,Sutherland2021}, quantum information processing \cite{BurdQuantum2019,xin2021RapidQuantum,Sutherland2021},  quantum thermodynamics \cite{Johannes2016HeatEngine}, quantum metrology \cite{SaNeto2022temperature}, etc. 
 The dynamics of the center of mass of an ion trapped by  periodic potential fields is described, analogously, by the solutions of a Quantum Harmonic Oscillator (QHO) with time-dependent frequency \cite{Paul1990traps,Baseia1993,Leibfried2003Ions,brown2011coupled,DingOscillatorTrappedIons2017,bayen2023effects}.  The dynamics of this oscillator has been well discussed in the literature using various methods and approaches \cite{husimi1953,Lewis1969oscillator,Janszky1986,Rhodes1989,Graham1987,Lo1990,Agarwal1991oscillator,Janszky1992,Kiss1994evolution,kiss1994time,tibaduiza2020e,tibaduiza2021time} and one of its important characteristics is the ability to generate  squeezed states according to frequency manipulation. The Squeezed states \cite{walls1983squeezed,scully1999quantum} are considered non-classical characteristic of a system and  allow several applications in quantum information theory  \cite{BurdQuantum2019,xin2021RapidQuantum,Sutherland2021}, quantum thermodynamics \cite{Deffner2008Nonequilibrium,Galve2009nonequilibrium,kosloff2017quantum,Manzano2018squeezed,Alexssandre2022unravelling, kim2022nonadiabaticity}, and quantum metrology \cite{SaNeto2022temperature}. 

 For an adiabatic manipulation, the frequency changes very slowly in time and forbids squeezing generation, while for a sudden manipulation, the frequency changes in a short interval of time, and allows squeezing  generation \cite{Janszky1986,Graham1987, Janszky1992,kiss1994time}. In fact, an adequate configuration of multiple sudden changes in frequency allows generating a greater degree of squeezing \cite{Janszky1992}. This technique was used to discuss the nonequilibrium thermodynamics of a harmonic oscillator \cite{Galve2009nonequilibrium} and experimentally implemented in ion trap systems for quantum information processing   
\cite{BurdQuantum2019,xin2021RapidQuantum}.  In between the adiabatic and sudden change, it is known that non-adiabatic manipulation of some particular frequencies generates squeezing \cite{Agarwal1991oscillator,Lo1990,tibaduiza2020e,tibaduiza2021time,Kiss1994evolution}, but to a lesser degree in comparison to the squeezing generated in the case of a sudden change.  The non-adiabatic dynamics of the oscillator can be determined through the non-adiabatic parameter, $Q^{*}$, defined by Husimi \cite{husimi1953} to describe the dynamics and transition probabilities of a quantum oscillator using a Gaussian wave function ansatz. The non-adiabaticity parameter depends   on the shape and manipulation of the frequency -- for an adiabatic manipulation,  $Q^{*}=1$, while for a non-adiabatic manipulation,  $Q^{*}>1$, always.    This parameter was considered in \cite{Deffner2008Nonequilibrium,Galve2009nonequilibrium} to analyze the non-equilibrium thermodynamics of the quantum oscillator.  Here there is a relationship between the non-adiabaticity parameter and the non-classical behavior of the system.  Analyzing this relationship is important for processes with continuous variable quantum systems,  {where non-classical states} can show some quantum advantage in a continuous time process. This may be useful in quantum information processes and quantum thermodynamics.
 
 In this article,  we show that there is a critical value non-adiabaticity parameter, defined by the initial state, that allows distinguishing the classical and non-classical behavior of the system. { Squeezed vacuum states} are known as being non-classical, therefore, the critical value of the non-adiabaticity parameter allows us to identify that the squeezing generated on the system is capable of reducing the uncertainty below  that of the vacuum.  The non-classical behavior of the system is given by the concept of P-representability  \cite{Glauber1963Coherent,englert2003tutorial,Alexssandre2022unravelling} -- given the system state $\thickhat{\rho}$ spanned in a coherent state basis $\{|\alpha\rangle\}$:
\begin{equation}
    \thickhat{\rho}=\int d^{2}\alpha P\left(\alpha, \alpha^{*}\right)\vert \alpha\rangle \langle \alpha \vert,
\end{equation}
    whenever the quasiprobability distribution $P(\alpha,\alpha^{*})$ is positive, the system state is said to be classical, while for a non-positive $P(\alpha,\alpha^{*})$ the system is in a non-classical state.  

Using the equations of motion in the Heisenberg picture, we find the relationship between the non-adiabatic dynamics and the non-classicality of the system. However, this procedure does not adequately show the evolution of squeezing in continuous time. Therefore, we consider the unitary evolution approach to propose a unitary time evolution operator analysis to describe the squeezing behavior. In this case we find that the squeezing parameter presents a critical value that differentiates the classical and non-classical behavior of the system. This type of result has already been discussed in particular in quantum thermodynamics for Gaussian heat engines \cite{Alexssandre2022unravelling}.

This work is organized as follows. In section \ref{sec:section2}, using the Heisenberg picture, we describe the non-adiabatic dynamics of the ion in terms of the non-adiabaticity parameter $Q^{*}$. Using the covariance matrix of the system, we define the classicality function to discuss the relationship between the non-adiabatic dynamics and the non-classical behavior of the trapped ion.  In section \ref{sec:section3}, we propose an ansatz of the time evolution operator to describe the behavior of squeezing during the evolution of the system. The time-dependent parameters that define the time evolution operator are only obtained using the Heisenberg equations of motion. In section \ref{sec:section4}, we analyze the relationship between non-adiabatic dynamics and the non-classical behavior of the trapped ion in a periodic electromagnetic potential field, preparing the initial state of the ion in
 a state of thermal equilibrium. In Section \ref{sec:section5} we present our conclusions and final remarks. 

%
%
%
\section{\label{sec:section2} Dynamics of a single trapped ion}
%
%
The dynamics of an ion trapped in an electromagnetic potential field is analogous to a QHO with time-dependent frequency. Therefore, the Hamiltonian of a single trapped ion can be defined as
\begin{equation}\label{eq.hamiltonianQHO}
    \thickhat{H}(t)=\frac{\thickhat{p}^{2}}{2m}+\frac{1}{2}mw^{2}(t) \thickhat{x}^2,
\end{equation}
where $m$ is the mass of the ion and $w(t)$  is the time-dependent frequency representing the manipulation of the electromagnetic potential field. $\thickhat{x}$ and $\thickhat{p}$ represent the position and momentum operators of the ion's center of mass, respectively.

 The Hamiltonian can be written in the form  $\thickhat{H}(t)=\hbar w(t)\left(\thickhat{b}^{\dagger}(t)\thickhat{b}(t)+\frac{1}{2}\right)$, where we define the time-dependent   operators in the Schrödinger picture, as   $\thickhat{b} (t)=\sqrt{\frac{m w(t)}{2\hbar}}\left(\thickhat{x}+i\frac{\thickhat{p}}{mw(t)} \right)$ and $  \thickhat{b}^{\dagger}(t)\equiv \left[ \thickhat{b}(t)\right]^{\dagger}$, therefore,  for each value of the parametrized frequency we have new   operators. For equal times, these operators satisfy the following commutation relations: $[\thickhat{b}(t),\thickhat{b}(t)]=0$, and $[\thickhat{b}(t),\thickhat{b}^{\dagger}(t)]=1$, while, the commutation relations in different time is   $[\thickhat{b}(t'),\thickhat{b}(t)]\neq 0$,  implying in  $[\thickhat{H}(t'), \thickhat{H}(t)]\neq 0$. For  $t=0$,   Eq. (\ref{eq.hamiltonianQHO}) is reduced to the time-independent Hamiltonian with initial frequency $w(0)=w_{0}$, that is, $\thickhat{H}(0)\equiv \hbar w_{0}\left(\thickhat{a}^{\dagger}\thickhat{a}+\frac{1}{2}\right)$. The relationship between annihilation (creation) operators $\thickhat{a}$ ($\thickhat{a}^{\dagger}$) of the free harmonic oscillator  and $\thickhat{b}(t)$($\thickhat{b}^{\dagger} (t)$) is determined by the Bogoliubov transformation,
\begin{equation}\label{eq.Bogoliubovtrans}
\begin{split}
\thickhat{b}^{\dagger}(t)&=\cosh(r_{a})\thickhat{a}^{\dagger} +\sinh(r_{a})\thickhat{a}=\thickhat{S}(r_{a})\thickhat{a}^{\dagger}\thickhat{S}^{\dagger}(r_{a}),\\ 
\thickhat{b}(t)&=\cosh(r_{a})\thickhat{a} +\sinh(r_{a})\thickhat{a}^{\dagger}=\thickhat{S}(r_{a})\thickhat{a}\thickhat{S}^{\dagger}(r_{a}),
\end{split}    
\end{equation}
where $e^{r_{a}}=\sqrt{\frac{w(t)}{w_{0}}}$,  $ \thickhat{S}(r_{a})= e^{\frac{r_{a}}{2}\left(\thickhat{a}^{\dagger 2}-\thickhat{a}^{2}\right)}$ is the squeezing operator, with squeezing parameter  $r_{a}$. The evolution of the system is described by the unitary time evolution operator    $\thickhat{U}(t,0)=\thickhat{T}\lbrace e^{-\frac{i}{\hbar}\int^{t}_{0} \thickhat{H}(t')dt'}\rbrace$, where $\thickhat{T}$ is the time ordering operator. The  operator, $\thickhat{U}(t,0)$, evolves the oscillator from an initial state $\thickhat{\rho}_{0}$ to a final state $\thickhat{\rho}(t)\equiv \thickhat{U}(t,0)\thickhat{\rho}_{0}\thickhat{U}^{\dagger}(t,0)$. 
 
Let us firstly consider the sudden approximation -- the frequency changes in a very short time interval in such a way that the system remains in the same  state when the manipulation occurs. Then, the time evolution operator is given by the identity, $\thickhat{U}_{sudd}(t,0) \equiv \thickhat{\mathcal{I}}$ \cite{bohm2012quantum,hecht2012quantum}. In this case, the squeezing effect is produced by $\thickhat{S}(r_{a})$ \cite{kiss1994time,Janszky1986,Graham1987,Janszky1992}, and describes the evolution of the system by jumping from an initial frequency $w_{0}$ to a final frequency $w_{1}>w_{0}$.  This type of manipulation was implemented to amplify coherent displacements   \cite{BurdQuantum2019}  and creating a squeezed state \cite{xin2021RapidQuantum}.

On the other hand, in the adiabatic approximation, as the frequency changes slowly, it is possible to show that the adiabatic time evolution operator, $\thickhat{U}_{ad}$, in fact, counteracts the squeezing effect produced by the Bogoliubov transformation.  Therefore, while the expected value of the Hamiltonian (Energy) of the system during the evolution will depend on the protocol employed for the frequency change, the expected value of the particle number operator remains constant through the evolution.
 
The dynamics of the system observables in an arbitrarilly non-adiabatic situation is analyzed using the Heisenberg picture (see Appendix \ref{apendiceA}). To reduce the mathematical expressions, we will consider $w(t)\equiv w$.  The equations of motion of the position, $\thickhat{x}_{H}$, and momentum, $\thickhat{p}_{H}$ result in   (we use the subscript $\lbrace H \rbrace$ to identify the operators in the Heisenberg picture) 
\begin{equation}\label{eq.xmotion} 
   \ddot{\thickhat{x}}_{H}+w^{2}\thickhat{x}_{H}=0.
\end{equation}
 The solution of (\ref{eq.xmotion}) is given by $ \thickhat{x}_{H}=u\thickhat{x}+v\thickhat{p}\quad  $ and $\quad \thickhat{p}_{H} = \bar{u} \thickhat{x} +\bar{v} \thickhat{p} $, where $u$, $v$, $\bar{u}\equiv m\dot{u}$ and  $\bar{v}\equiv m\dot{v}$ are real time-dependent functions, which must satisfy the differential equations 
 \begin{equation}\label{eq.uvMathieu}
     \ddot{u}+w^{2}u=0,\quad \text{and} \quad \ddot{\bar{v}}+w^{2}\bar{v}=0, 
 \end{equation}
 with the  initial conditions $u(0)=1$, $\dot{u}=0$ and $\bar{v}(0)=0$, $\dot{\bar{v}}(0)=1$, respectively. For the commutation relation between $\thickhat{x}_{H}$ and $\thickhat{p}_{H}$ to remain physical, it must be true that    
 \begin{equation}\label{eq.condition1}
 u\dot{\bar{v}}-\dot{u}\bar{v}=1.
 \end{equation}
An arbitrary frequency $w$ is physically valid if the solutions $u$ and $\bar{v}$ of (\ref{eq.uvMathieu}) are real functions and satisfy condition (\ref{eq.condition1}).  
 
 For the dynamics of the system,  we define the Lagrangian operator $\thickhat{\mathcal{L}}_{H}= \frac{1}{2m}\thickhat{p}^{2}_{H}-\frac{1}{2}mw^{2}\thickhat{x}^{2}_{H}$ and the position-momentum correlation operator, $\thickhat{\mathcal{C}}o_{H} =\frac{w}{2}\left(\thickhat{x}_{H}\thickhat{p}_{H}+\thickhat{p}_{H}\thickhat{x}_{H}\right)$, these operators, together with the Hamiltonian  $\thickhat{H}_{H} $, satisfy a Lie algebra \cite{hoffmann2013casimir,kosloff2017quantum}. 
 
 The mean value of the Hamiltonian, Lagrangian   and   position-momentum correlation operator  of the system are   related to three parameters, $Q^{*} $,  $Q^{*}_{1} $ and $Q^{*}_{2} $, respectively, that is (see Appendix \ref{apendiceA})
\begin{equation}\label{eq.nonadiabaticity}
\begin{split}
    Q^{*} &= \frac{1}{2w_{0}w}\left[w^{2}_{0}\left(\dot{\bar{v}}^{2} +w^{2}\bar{v}^{2}\right)+w^{2}u^{2}+\dot{u}^{2}\right],\\
    Q^{*}_{1}&=\frac{1}{2w_{0}w}\left[w^{2}_{0}\left(\dot{\bar{v}}^{2}-w^{2}\bar{v}^{2}\right)    +  \dot{u}^{2}-w^{2}u^{2}\right],\\
Q^{*}_{2} &= \frac{1}{2w_{0}w}\left[2w u\dot{u} +2w w^{2}_{0} \bar{v}\dot{\bar{v}} \right].
\end{split}
\end{equation}

The parameter{\footnote{It is worth to mention that this parameter has no relation to the Husimi quasi-probability distribution $Q=\frac{1}{\pi}\langle \alpha \vert \thickhat{\rho}\vert \alpha\rangle$ previously introduced  in \cite{husimi1}}}  $Q^{*}$ represents the non-adiabaticity parameter \cite{husimi1953,Deffner2008Nonequilibrium, Galve2009nonequilibrium}, and describes the behavior of the system according to the protocol considered for the frequency $w$. For an adiabatic manipulation, the frequency changes slowly over time, the non-adiabaticity parameter is $Q^{*}=1$. This means that the system evolves following the adiabatic theorem of Quantum Mechanics and, therefore, there are no transitions occurring in the energy levels of the system.  For a certain energy level, the average value of the number of particles must remain constant. While for a non-adiabatic process, $Q^{*}>1$, there are transitions between the different energy levels of the system, therefore, the average value of the number of particles will change during the evolution of the system. 
 
From Eq. (\ref{eq.nonadiabaticity}) and using condition (\ref{eq.condition1}), we find that 
\begin{equation}\label{eq.relationQQ}
    {Q^{*}}^{2}-\left({Q^{*}_{1}}^{2} +{Q^{*}_{2}}^{2}\right)=1.
\end{equation}

The average values of the operators $\thickhat{\mathcal{L}}_{H}$ and $\thickhat{\mathcal{C}}o_{H}$ contribute to the coherence of the system. In fact, the covariance matrix representing the system Gaussian state can be written in terms of $\thickhat{H}_{H} $, $\thickhat{\mathcal{L}}_{H}$, and $\thickhat{\mathcal{C}}o_{H}$ as
\begin{equation}
    \bm{V}=\frac{1}{\hbar w}\left(\begin{array}{c c}
         \langle \thickhat{H}_{H}\rangle  & -\langle \thickhat{\mathcal{L}}_{H}\rangle +i \langle\thickhat{\mathcal{C}}o_{H}\rangle \\
         -\langle \thickhat{\mathcal{L}}_{H}\rangle -i \langle \thickhat{\mathcal{C}}o_{H}\rangle& \langle \thickhat{H}_{H}\rangle
    \end{array}\right),
\end{equation}
which is always positive. 

From Eqs.  (\ref{eq.relationQQ}) the relationship between non-adiabatic dynamics and the coherence of the system is observed. If the process is adiabatic ($Q^{*}=1$), it is expected that the system  will present classical behavior, therefore, it will not present coherence. While as the degree of non-adiabaticity increases ($Q^{*}>1$), it is expected that the system will present a greater degree of coherence, and therefore, the oscillator will present a non-classical behavior.  

The classicality criterion of the states of the system is determined by the condition    $\bm{V}-\frac{1}{2}\bm{I}\ge 0$, where $\bm{I}$ is the $ 2\times 2$ identity \cite{englert2003tutorial,MCO1}, which rewritten explicitly, leads to the classicality function \cite{Alexssandre2022unravelling} as 
\begin{equation}\label{eq.clasicalidade} \mathcal{C}(\bar{n},Q^{*})=\left(\bar{n}+\frac{1}{2}\right) \left[ Q^{*}-\sqrt{{Q^{*}}^{2} -1 } \right]-\frac{1}{2},
\end{equation}
where the relation of Eq.  (\ref{eq.relationQQ}) was applied. The classicality function helps us to identify the classical and non-classical behavior of the system.   The states of the trapped ion are classical for $\mathcal{C}(\bar{n},Q^{*})>0$, while for $\mathcal{C}(\bar{n},Q^{*})<0$, the states are non-classical. 

From Eq. (\ref{eq.clasicalidade}), the function $\mathcal{C}(\bar{n},Q^{*})$ depends on the behavior of two parameters: the value of the mean value of the number operator in an initial state and the non-adiabatic manipulation of the potential.  So when the process is adiabatic, $Q^{*}=1$,  the state of the system will be classical  regardless of $\bar{n}$.  However, in a non-adiabatic process, the parameter $Q^{*}$ must reach a critical value in order to distinguish between classical and non-classical states.

For $\mathcal{C}(\bar{n},Q^{*})=0$,  we find the critical value of the non-adiabaticity parameter, $Q^{*}_{c}$, that is 
 \begin{equation}
     Q^{*}_{c}\equiv \frac{\left(\bar{n}+\frac{1}{2}\right)^{2} +\frac{1}{4}}{\bar{n}+\frac{1}{2}}.
 \end{equation}

\begin{figure} [ht]
	\centering
	\includegraphics[width=1.0\linewidth]{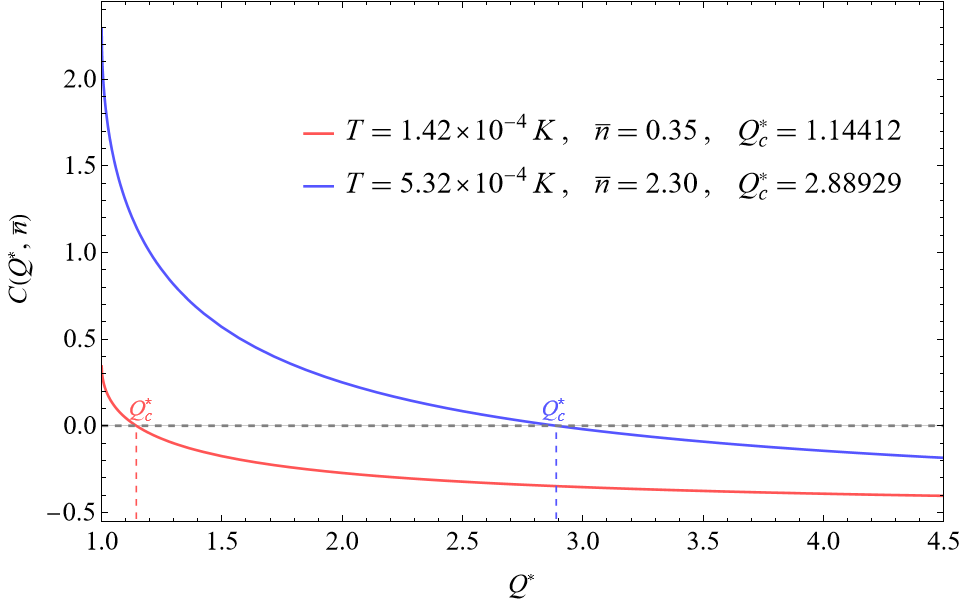}
	\caption{\justifying  Experimental values of $\bar{n}$ with their respective temperatures  in ion trap systems \cite{brown2011coupled,SaNeto2022temperature}. To obtain non-classical states, it is necessary for the parameter $Q^{*}$  exceeds its critical value for the corresponding value of $\bar{n}$. As the value of $\bar{n}$ increases, the value of $Q^{*}_{c}$ will increase.
    } 
	\label{fig.classicality}
\end{figure}

 For $Q^{*}<Q^{*}_{c}$  the classicality function is positive, therefore, the state will be classical, while for $Q^{*}>Q^{*}_{c}$, the classicality function will be negative, consequently, the state will be non-classical. Figure \ref{fig.classicality} shows the relationship between the classicality function and the non-adiabaticity parameter. The critical value of the non-adiabaticity parameter acts as an identifier of the non-classicality of the trapped ion.   The greater the value of $\bar{n}$, the greater the critical value, therefore, it will be necessary to properly manipulate the potential    to achieve non-classical behavior.

From the classicality function (\ref{eq.clasicalidade}), we can observe that a non-adiabatic manipulation does not necessarily generate non-classical states ($Q^{*}<Q^{*}_{c}$), while non-classicality is a sufficient condition to affirm that the dynamics  of the trapped ion is non-adiabatic ($Q^{*}>Q^{*}_{c}$).  

  Squeezing is a non-classical characteristic of the system, however, we show that this will depend on the critical value of the degree of non-adiabaticity, meaning that there is a critical value of the squeezing parameter related to $Q^{*}_{c}$, therefore, when the squeezing parameter exceeds its critical value, the uncertainty of one of the quadratures will be lower than that of the vacuum state. In the case of not exceeding its critical value, there is a "squeezing" effect on the state of the oscillator but the uncertainty in one of its quadratures will not reach the uncertainty of the vacuum. It is worth noting that the classicality function and the critical value of the squeezing parameter was discussed in Otto cycle using the single mode oscillator as working substance \cite{Alexssandre2022unravelling}, and therefore, there are direct implications of the non-adiabaticity in the efficiency of quantum thermal engines. 
 
%
\section{\label{sec:section3} Time Evolution Operator}
%
%
%
In the previous section we showed the non-adiabatic dynamics of the trapped ion and the squeezing effect could be determined by analyzing the variance of the quadratures related to the position and momentum operators. However, this does not allow us to identify the evolution of the squeezing phase and squeezing parameter. Therefore, we propose a second approach to observe the squeezing behavior by describing the time evolution operator of the trapped ion.
 
Some algebraic methods were previously applied to find the time evolution operator of a harmonic oscillator with time-dependent frequency \cite{Rhodes1989,Kiss1994evolution}. Furthermore, Lie algebra methods were applied to relate the time evolution operator in terms of the squeezing operator \cite{ Truax1985BCH, rezek2006irreversible,naudts2011bch,martinez2020newBCH,tibaduiza2021time,tibaduiza2020e}. {{ We propose a time evolution operator { of the system as} $\bar{U}(t,0)$, such that if $\thickhat{\mathcal{O}}$ is an operator in the Schrodinger picture, then { in the Heisenberg picture} its  time evolution  is defined as $\thickhat{\mathcal{O}}_{t} {\equiv}\bar{U}^{\dagger}(t,0)\thickhat{\mathcal{O}} \bar{U}(t,0)$.  From  equations (\ref{eq.xpsolution01}),  we can identify that the time evolution of these operators using $\bar{U}(t,0)$ will have the same form. For simplicity, we use the creation (annihilation) operators $\thickhat{a}^{\dagger}$($\thickhat{a}$), so the time evolution of these operators are defined as }

\begin{equation}\label{eq.unitaad01}
\begin{split}
    \thickhat{a}_{t}&\equiv \thickhat{U}^{\dagger}(t,0) \thickhat{a}\thickhat{U}(t,0)= f  \thickhat{a}+g \thickhat{a}^{\dagger},\\
    \thickhat{a}^{\dagger}_{t}& \equiv \thickhat{U}^{\dagger}(t,0) \thickhat{a}^{\dagger}\thickhat{U}(t,0)= f^{*} \thickhat{a}^{\dagger}+g^{*} \thickhat{a},\\   
\end{split}
\end{equation}
where the time dependent functions $f $, $g $ $\in \mathbb{C}$. 
Using equation (\ref{eq.unitaad01}), we write the time evolution of the position and momentum operators as  $\thickhat{x}_{t}=\sqrt{\frac{\hbar}{2mw_{0}}} \left(\thickhat{a}^{\dagger}_{t}+\thickhat{a}_{t}\right)$ and $\thickhat{p}_{t}=i\sqrt{\frac{\hbar m w_{0}}{2}} \left(\thickhat{a}^{\dagger}_{t}-\thickhat{a}_{t}\right)$, that is
\begin{equation}\label{eq.xpunitario}
\begin{split}
    \thickhat{x}_{t} &=  \sqrt{\frac{\hbar}{2mw_{0}}}  \left[\left( f^{*}  + g \right)\thickhat{a}^{\dagger}  + \left( g^{*} +f \right)\thickhat{a}  \right],\\
    \thickhat{p}_{t} &=  i \sqrt{\frac{\hbar m w_{0}}{2}} \left[\left( f^{*}  - g \right)\thickhat{a}^{\dagger}  - \left( f - g^{*} \right)\thickhat{a} \right].
\end{split}
\end{equation}

The operators $\thickhat{x}_{t}$ and $\thickhat{p}_{t}$ must be equal to the operators $\thickhat{x}_{H}$ and $\thickhat{p}_{H}$, respectively. Therefore, we have that
 \begin{equation}\label{eq.fgfunction}
\begin{split}
    f&=\frac{1}{2}\left(u +   \dot{\bar{v}} \right) - \frac{i}{2} \left( w_{0}\bar{v} -\frac{\dot{u}}{w_{0}} \right),\\
    g&=\frac{1}{2}\left(u - \dot{\bar{v}} \right) +\frac{i}{2} \left(w_{0}\bar{v} +\frac{\dot{u}}{w_{0}} \right).
\end{split}
\end{equation}

In addition, {the operators $\thickhat{x}_{t}$ and $\thickhat{p}_{t}$ must satisfy the commutation relation}  $\left[\thickhat{x}_{t},\thickhat{p}_{t}\right]=i\hbar$,  {therefore, the functions $f$ and $g$ must satisfy the following condition}
\begin{equation}\label{eq.condition02}
    \vert  f\vert^{2} -\vert g\vert^{2}=1,
\end{equation} 
where the condition of Eq. (\ref{eq.condition1}) was applied.  
 
So, it is possible to define a time evolution operator in such a way that applied to $\thickhat{a}$($\thickhat{a}^{\dagger}$)  it has the form of Eq. (\ref{eq.unitaad01}) and that condition  (\ref{eq.condition02})  is satisfied. We propose an   ansatz of the time evolution operator of the form  
\begin{equation}\label{eq.opunitario1}
    \bar{U}(t,0)=\thickhat{S}(\xi) \thickhat{R}(\gamma)
\end{equation}
where $\thickhat{S}(\xi)$ and $\thickhat{R}(\gamma )$ are the squeezing and rotation operators, respectively. These unitary operators are defined as
\begin{equation}\label{eq.SqueezeRotation}
\begin{split}     
    \thickhat{S}(\xi)&= e^{\frac{1}{2}\left( \xi^{*} \thickhat{a}^{2}-\xi\thickhat{a}^{\dagger 2}\right)},\\
    \thickhat{R}(\gamma)&= e^{-i \gamma\left(\thickhat{a}^{\dagger}\thickhat{a} +\frac{1}{2}\right)},
\end{split}
\end{equation}
where $\xi= r e^{i\theta}$ and   $r$, $\theta$ and $\gamma$ are time-dependent variables. Applying the evolution operator on the position and momentum operators,  we have  
\begin{eqnarray}\label{eq.xt}
\thickhat{x}_{t} =&&\sqrt{\frac{\hbar}{2mw_{0}}}  \left[ \left( \cosh(r) - e^{-i\theta } \sinh(r)\right)
\thickhat{a}e^{-i\gamma} \right. \nonumber \\
 & &+ \left. \left(\cosh(r) -e^{i\theta }\sinh(r)   \right)\thickhat{a}^{\dagger}e^{i\gamma}  \right].
\end{eqnarray}
\begin{eqnarray}\label{eq.pt}
\thickhat{p}_{t} =&& i \sqrt{\frac{\hbar m w_{0}}{2}} \left[ \left( \cosh(r) + e^{i\theta } \sinh(r)\right) 
\thickhat{a}^{\dagger}e^{i\gamma} \right. \nonumber \\
 & &- \left. \left(\cosh(r) + e^{-i\theta} \sinh(r)   \right)\thickhat{a}e^{-i\gamma}  \right]. 
\end{eqnarray}
The time dependent parameters of $\bar{U}$ are obtained by equating equations 
 (\ref{eq.xt}) and (\ref{eq.pt}) with (\ref{eq.xpunitario}), that is 
\begin{equation}\label{eq.parameters}
\begin{split}
    r&=\arccosh\left( \vert f \vert    \right),\\
       \theta &= \varphi_{f}  +\varphi_{g} \pm (2n +1)\pi,\\
        \gamma &= -\varphi_{f}.
\end{split}
\end{equation}
where  $n= 0, 1, 2,\cdots$,  $ \varphi_{f}= arg(f)$ and $ \varphi_{g}= arg(g)$ are the arguments of $f$ and $g$, respectively. {Eqs. (\ref{eq.parameters}) and (\ref{eq.fgfunction}) show the relationship between the squeezing and rotation parameters with the solutions of the Heisenberg equations of motion.}

 The evolution of the state of the system is determined by the temporal evolution operator. Considering the initial state of the system as $\thickhat{\rho}_{0}$, the final state of the system is described by $\thickhat{\rho}(t)= \bar{U}(t,0) \thickhat{\rho}_{0}\bar{U}^{\dagger}(t,0)$, that is, 
\begin{equation}
    \thickhat{\rho}(t)=\thickhat{S}(\xi) \thickhat{R}(\gamma) \thickhat{\rho}_{0}  \thickhat{R}^{\dagger}(\gamma) \thickhat{S}^{\dagger}(\xi).
\end{equation}
%
%
%
%
%

{Now, using the operators $\thickhat{x}_{t}$ and $\thickhat{p}_{t} $, we analyze the dynamics of the trapped ion in terms of the parameters of the time evolution operator. In a manner analogous to the Heisenberg picture approach,  we construct the Hamiltonian ($\thickhat{H}_{t}$), Lagrangian ($\thickhat{\mathcal{L}}_{t}$) and the position-momentum correlation ($\thickhat{\mathcal{C}}o_{t}$) operators. Then, the expected values of these operators have the same form as Eq. (\ref{eq.HamiltQ01}), where the parameters $Q^{*}$, $Q^{*}_{2}$ and $Q^{*}_{3}$ will be of the form}  
\begin{equation}\label{eq.QsUoperator}
 \begin{split}
Q^{*}&=\cosh(2r)\cosh(2r_{a})-\sinh(2r)\sinh(2r_{a})\cos(\theta),\\ 
Q^{*}_{1}&=\sinh(2r)\cosh(2r_{a})\cos(\theta)-\cosh(2r)\sinh(2r_{a}),\\ 
Q^{*}_{2}&=\sinh(2r)\sin(\theta),\\ 
 \end{split}
\end{equation}
where $Q^{*}$ is the non-adiabaticity parameter obtained using the time-evolution approach. These  parameters satisfy the condition of Eq. (\ref{eq.relationQQ}).

{ We find that for an adiabatic process, $Q^{*}=1$ and, is obtained when $\theta \approx 2 n \pi$ (with  $n=0,1,2,\cdots$) and $r=r_{a}$, while the parameters  $Q^{*}_{1}$ and  $Q^{*}_{2}$ are null. Therefore, we identify the time evolution operator as an adiabatic evolution operator $\bar{U}_{ad}$ where the squeezing effect given by $r=r_{a}$ counteracts the squeezing effect produced by the Bogoliubov transformation of Eq. (\ref{eq.Bogoliubovtrans}).}  

\section{\label{sec:section4} Protocol}

Initially, we consider that the initial state of the ion is prepared in thermal equilibrium with a thermal bath with inverse temperature  $\beta=\frac{1}{k_{B}T}$, where $k_{B}$ is the Boltzmann constant and $T$ is the temperature of the bath. In the coherent state representation, the initial state of the system is
\begin{equation}\label{eq.estadoinicial}
	\thickhat{\rho}_{0}=\frac{1}{\pi \bar{n}}\int e^{-\frac{\vert \alpha \vert^{2}}{\bar{n}} }\vert \alpha \rangle\langle \alpha \vert d^{2}\alpha,
\end{equation}
where $\bar{n} =\left[e^{\beta \hbar w_{0}}-1\right]^{-1} $ represents the number of thermal bosons. So the initial energy of the system is $E_{0}=\frac{\hbar w_{0}}{2}\coth\left(\frac{\beta \hbar w_{0}}{2}\right)$.

We consider a  manipulated potential of the form  $ w^{2}=a-2q   \cos(2\tau) $, where $\tau=\phi t$, in such a way that it goes from an initial value $w_{0}$   to a final value $w_{1}>w_{0}$ in a finite time.  Then, from Eq. (\ref{eq.uvMathieu}), the differential equations of $u$ and $v$ describe a Mathieu's equation.  We analyze the results using the time scale $\tau=\phi t$. Rewriting the Mathieu equation of $u$ and $\nu$, we find that the parameters $\bar{a}=\frac{a}{\phi^{2.}}$ and $\bar{q}=\frac{q}{\phi^{2}}$ describe the regions of stability and instability, that is, the solutions can be stable, periodic and unstable.  For a stable solution, its amplitude must decrease or approach a constant value. Conversely, the amplitude of an unstable solution will increase unlimitedly \cite{jordan2007nonlinear,Jazar2021perturbation,kovacic2018mathieu}. In the case of a periodic solution, the amplitude will be constant.   For the numerical analysis of the protocol we consider experimental data of related to vibrational degrees of freedom of trapped ions, where the average number of thermal photons is $\bar{n}=0.35$ at temperature $T=1.42 \times 10^{-4} \si{\kelvin}$ and with initial frequency $\frac{w_{0}}{2\pi}=4 \si{\mega \hertz}$ accordingly to \cite{brown2011coupled,SaNeto2022temperature}.
      
Initially the  state of the ion is set up in a classical (theremal) state. As the parametric frequency changes over time, the system will present classical or non-classical behavior during its evolution depending on it regime. In the Mathieu stable region, for suitable values of the parametric frequency, we find that the dynamics of the ion can be quasi-adiabatic. As shown in Fig.\ref{fig.pontoEstavelpequeno}, the parameter $Q^{*}$ varies in the order of $10^{-3}$, therefore, it will not exceed its corresponding critical value $Q^{*}_{c}$. Using the time evolution operator approach, we found that, during the evolution, the squeezing phase tends to approach values of $2n\pi$ ($n=0, 1, 2,\cdots$), that is, $\cos(\theta (\tau)) \approx 1$, while the squeezing parameter (in Figs. \ref{fig.pontoEstavelpequeno} (c),(d) ) approache the value of the squeezing parameter of the Bogoliubov transformation, that is, $r\approx r_{a}$.
\begin{figure} [ht] 
	\centering
	\includegraphics[width=1.0\linewidth]{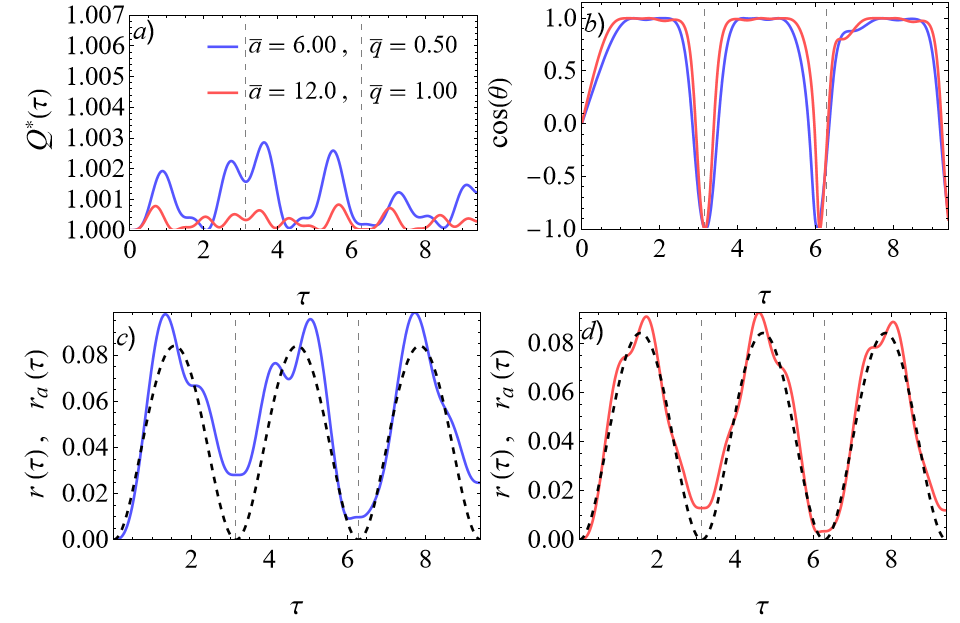}
	\caption{\justifying Quasi-adiabatic dynamics in the Mathieu stability region. The segmented lines correspond to the period of the frequency. a) Evolution of the non-adiabatic parameter.  b) Behavior of the squeezing phase. c) For $\lbrace\bar{a},\bar{q}\rbrace =\lbrace 6.00, 0.50\rbrace $  and d) $\lbrace\bar{a},\bar{q}\rbrace=\lbrace 12.0, 1.00 \rbrace$,   the squeezing parameter approaches the Bogoliubov squeezing parameter, $r\approx r_{a}$.  } 
	\label{fig.pontoEstavelpequeno}
\end{figure}   

Within the stable region, we find that the ion presents non-classical characteristics. During the evolution, we can observe that the non-adiabaticity parameter manages to exceed its critical value. However, due to the oscillatory behavior of the solutions of Mathieu's equation, and consequently,  of the parameter $Q^{*}$, the system will present a non-classical behavior for a certain period of time and will return to a classical behavior, as shown in Fig. \ref{fig.pontoEstavel}.  If the parameters $\lbrace\bar{a}, \bar{q}\rbrace$ are very close to the instability region, there is a greater period of oscillation of the parameter $Q^{*}$, and consequently, of the classicality function. 

\begin{figure} [ht] 
	\centering
	\includegraphics[width=1.0\linewidth]{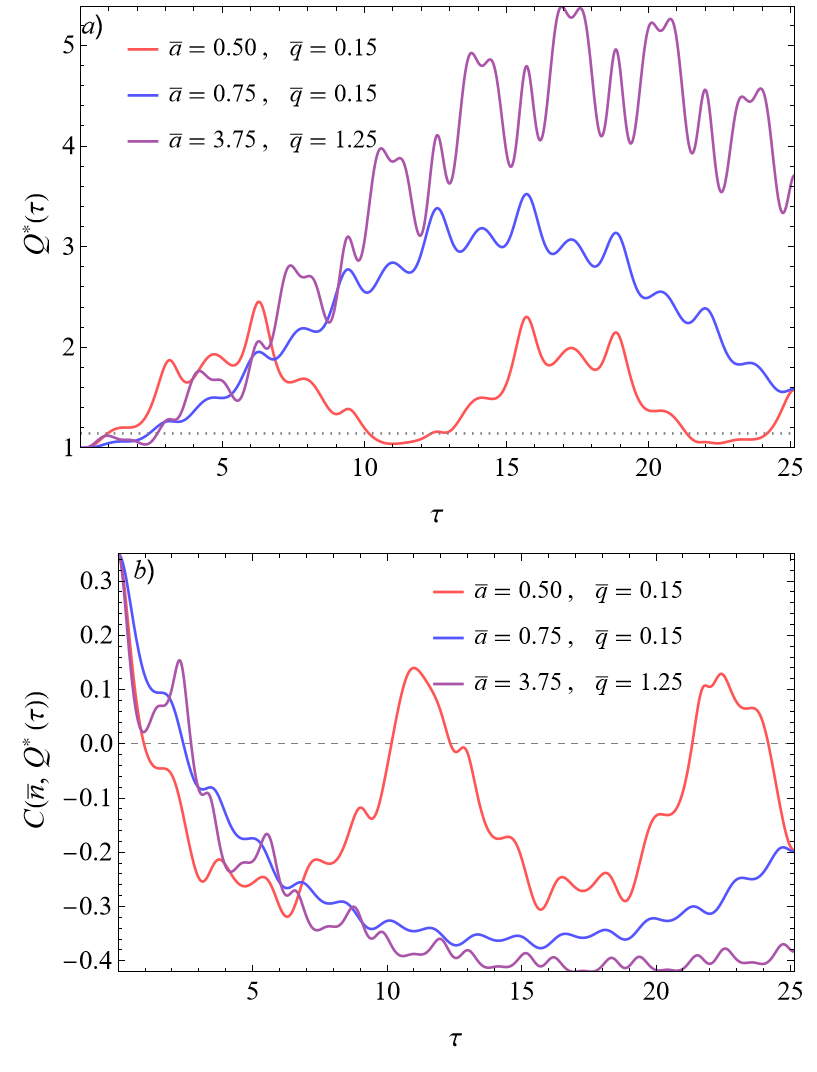}
	\caption{\justifying  Non-adiabatic behavior in the Mathieu stability region. a) The non-adiabaticity parameter presents an oscillatory behavior. This parameter manages to exceed its corresponding critical value, in such a way that the dynamics is non-classical during the time in which $Q^{*}>Q^{*}_{c}$ . b) Oscillatory behavior of the classicality function. } 
	\label{fig.pontoEstavel}
\end{figure} 

In the region of instability, the solutions of the Mathieu equation grow in an unlimited way, and consequently, the non-adiabaticity parameter $Q^{*}$  and the squeezing parameter will also grow as $\tau$ increases. Additionally, the classicity function tends to decay to its minimum value, this shows that the ion will present a greater degree of non-classicality.  In this region, Fig. \ref{fig.pontoInstavel},  the non-adiabaticity parameter tends to easily exceed its critical value, and therefore, the system is more prone to exhibit non-classical behavior.   Another way to observe the non-classicality of the oscillator is shown in the Fig. \ref{fig.pontoInstavel}(d), in which we have the curves $\langle n_{H}(\tau)\rangle  = \left(\bar{n} +\frac{1}{2} \right)Q^{*} -\frac{1}{2}  $ and $\vert m_{H}(\tau)\vert  = \left(\bar{n}+\frac{1}{2}\right) \sqrt{ Q^{ * 2} -1}   $  corresponding to the diagonal and off-diagonal components of the covariance matrix.  The blue region corresponds to the moment when  the state of the system is classical, that is, for $\mathcal{C}(\bar{n},Q^{*})>0$, while the red region corresponds  to non-classical behavior, that is,  $\mathcal{C}(\bar{n},Q^{*})<0$. The threshold point represents $\mathcal{C}(\bar{n},Q^{*})=0$, and therefore, it allows us to distinguish the classical and non-classical behavior of the oscillator. 
 \begin{figure} [ht!] 
	\centering
	\includegraphics[width=.82\linewidth]{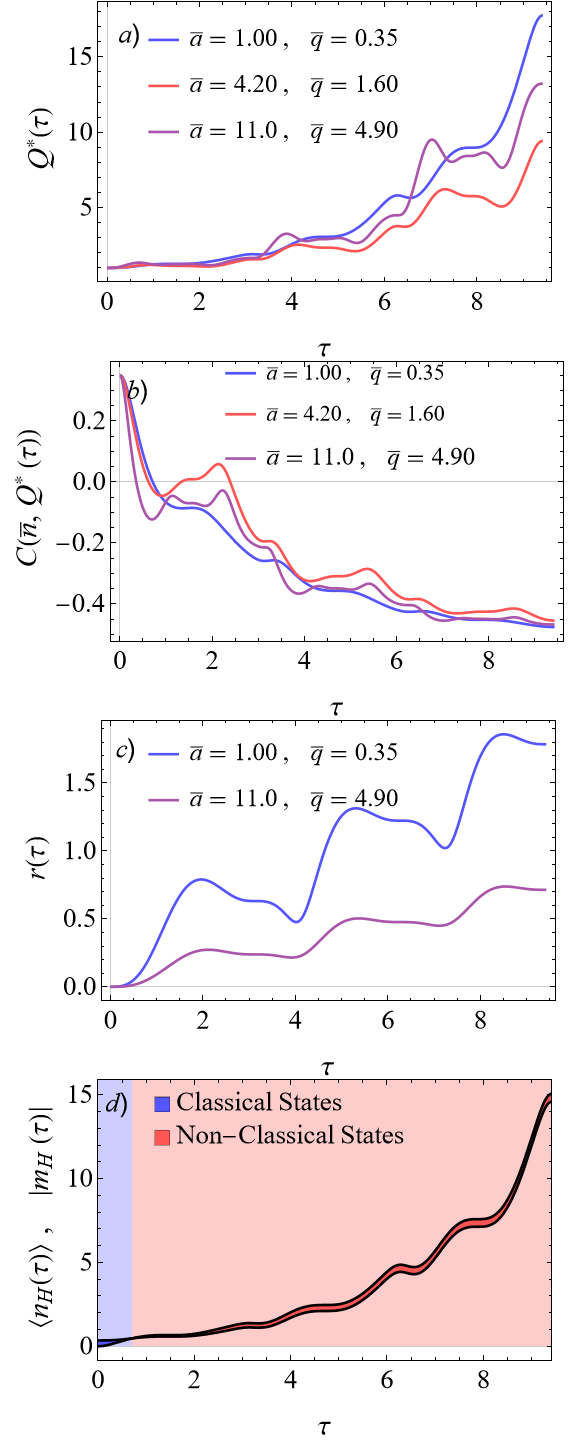}
	\caption{\justifying  Non-adiabatic behavior in the Mathieu     instability region. a) The non-adiabaticity parameter grows unlimitedly, therefore, it manages to exceed its corresponding critical value. b) The classicality function presents a more notable decay due to the increase in the non-adiabaticity parameter. c) Behavior of the squeezing parameter. d) Another way of discussing the non-classicality of the system, the blue region corresponds to the non-classical states for $\langle n_{H}(\tau)\rangle > \vert m_{H}(\tau)\vert$, while the red region corresponds to the non-classical states when $\langle n_{H}(\tau)\rangle < \vert m_{H}(\tau)\vert$. The point of intersection corresponds for $\mathcal{C}(\bar{n},Q^{*})=0$. } 
	\label{fig.pontoInstavel}
\end{figure} 

Considering the initial state in thermal equilibrium defined as $\thickhat{\rho}_{0}\equiv\frac{1}{\mathcal{Z}} e^{-\beta \thickhat{H}_{0}}$,  equivalent to  (\ref{eq.estadoinicial}), where $\thickhat{H}_{0}=\hbar w_{0}\left(\thickhat{a}^{\dagger}\thickhat{a}+\frac{1}{2}\right)$ and $\mathcal{Z}\equiv Tr\left[e^{-\beta \thickhat{H}_{0}} \right]$, then the state of the system evolves to a final state described by  $\thickhat{\rho}(t)= \bar{U}(t,0) \thickhat{\rho}_{0}\bar{U}^{\dagger}(t,0)$.

We know that when a squeezing operator is applied to a thermal state, the final state results in a squeezed thermal state \cite{Manzano2018squeezed}. In this case, the states of the system evolve in such a way that the final state can result in any state or in a squeezed thermal state, this is determined from the relationship of non-classicality and non-adiabaticity. So, we can say that the final state is a squeezed thermal state  in the regions that satisfy the condition $Q^{*}>Q^{*}_{c}$. This squeezed thermal state is
\begin{equation}
    \thickhat{\rho}(t)= \frac{e^{-\beta_{s} \left(\thickhat{H}_{0} - \mu \thickhat{A}_{s}\right)}}{\tilde{\mathcal{Z}}}.
\end{equation}
where the partition function is $\tilde{\mathcal{Z}}\equiv Tr\left[ e^{-\beta_{s} \left(\thickhat{H}_{0} - \mu \thickhat{A}_{s}\right)}\right]$, 
  the generalized inverse temperature $\beta_{s}=\beta \cosh(2r)$,  the chemical-like potential  $ \mu= \tanh(2r)$, $r$ and $\theta$ are defined from equation (\ref{eq.parameters}). The operator $\thickhat{A}_{s}$ is defined as
\begin{equation}
\thickhat{A}_{s}=-\frac{\hbar w_{0}}{2} \left(\thickhat{a}^{2}e^{-i \theta} + \thickhat{a}^{\dagger 2} e^{i\theta}\right)
\end{equation}

%
%
 
%
%
\section{\label{sec:section5} Discussions and conclusions}
%
%
%
%
{The nonadiabatic dynamics of a single  trapped ion without noise were described using the Heisenberg picture approach and the unitary time evolution operator approach. From the Heisenberg picture we deduce that the non-adiabiticity parameter depends on the solutions of the Heisenberg equations of motion. Therefore, $Q^{*}$ depends on the shape and   manipulation of frequency. Furthermore, the properties of the covariance matrix and P-representability show that there is a relationship between the non-adiabiticity and the non-classicality of the system. One of the non-classical features of the system is the degree of squeezing that can be generated during evolution. So, to analyze the evolution of squeezing, we propose a time-evolution operator ansatz, $\bar{U}= \thickhat{S}(\xi) \thickhat{R}(\gamma)$, such that the time-dependent parameters are related to the solutions of the Heisenberg equations of motion.  

The results obtained show that during a non-adiabatic evolution there are regions where the state of the system is classical or non-classical, in this case, these regions can be identified by the critical value of the non-adiabatic parameter, $Q^{*}_{c}$.  This result is important because it allows us to identify regions in which a non-classical state can be obtained in an adequate manner, because a necessary condition for this is that the non-adiabatic parameter exceeds its critical value, $Q^{*}>Q^{*}_{c}$.  Furthermore, discuss the behavior of the degree of squeezing   in a non-adiabatic manipulation situation is important for quantum information processes and quantum thermodynamics using a system of continuous variables. Within quantum thermodynamics, these results can be used to discuss the influence of non-adiabaticity  in quantum thermal engines and in resource theory. Lastly, we believe that the present results could be further extended to explore alternative trap geometries, such as those investigated in \cite{palmero}, particularly the fast rotation of an ion in a tightly confined rotating trap. This extension may offer deeper insights into the interplay between non-adiabatic dynamics and non-classical behavior in continuous variable quantum systems.}

\begin{acknowledgments}
 
\end{acknowledgments}
This study was financed in part by the Coordenação de Aperfeiçoamento de Pessoal de Nível Superior – Brasil (CAPES) – Finance Code 001, and by CNPq. \\
 
 \appendix
\section{Heisenberg Picture}\label{apendiceA}
The Hamiltonian of a single trapped ion is analogous to the Hamiltonian of a QHO with time-dependent frequency, i.e.
\begin{equation} 
    \thickhat{H}(t)=\frac{\thickhat{p}^{2}}{2m}+\frac{1}{2}mw^{2}(t) \thickhat{x}^2.
\end{equation}
In the Heisenberg picture, from the equations of motion of the operators $\thickhat{x} $ and $\thickhat{p}$, we obtain the differential equation of an oscillator with time-dependent frequency 
\begin{equation}\label{eq.xmotion01}
   \ddot{\thickhat{x}}_{H}+w^{2}\thickhat{x}_{H}=0
\end{equation}
where $w(t)\equiv w$. We propose a solution of the form 
\begin{equation}\label{eq.xpsolution01}
\thickhat{x}_{H}=u\thickhat{x}+v\thickhat{p}\quad \text{and}\quad \thickhat{p}_{H} = \bar{u} \thickhat{x} +\bar{v} \thickhat{p}, 
\end{equation}
 where $u$, $v$, $\bar{u}=m\dot{u}$ and  $\bar{v}=m\dot{v}$ are real time-dependent functions, and these functions are solutions  of the differential equations
 \begin{equation}\label{eq.uvMathieu01}
     \ddot{u}+w^{2}u=0,\quad \text{and} \quad \ddot{\bar{v}}+w^{2}\bar{v}=0, 
 \end{equation}
 with the  initial conditions $u(0)=1$, $\dot{u}=0$ and $\bar{v}(0)=0$, $\dot{\bar{v}}(0)=1$, respectively. For the commutation relation between $\thickhat{x}_{H}$ and $\thickhat{p}_{H}$ to be true, it must be true that    
 \begin{equation}\label{eq.condition001}
 u\dot{\bar{v}}-\dot{u}\bar{v}=1.
 \end{equation}
 
For the dynamics of the system,  we define the Lagrangian operator $\thickhat{\mathcal{L}}$ and the position-momentum correlation operator $\thickhat{\mathcal{C}}o$ in the Heisenberg picture,
\begin{equation}
\begin{split}
    \thickhat{\mathcal{L}}_{H} &=\frac{1}{2m}\thickhat{p}^{2}_{H}-\frac{1}{2}mw^{2}\thickhat{x}^{2}_{H},\\
    \thickhat{\mathcal{C}}o_{H} &=\frac{w}{2}\left(\thickhat{x}_{H}\thickhat{p}_{H}+\thickhat{p}_{H}\thickhat{x}_{H}\right),
\end{split}
\end{equation}
 
In this work, we consider that the ion is prepared in an initial state $\rho_{0}$, in such a way that the position and momentum operators have mean zero,  that is, $\langle \thickhat{x}\rangle= 0$ and $\langle \thickhat{p}\rangle=0$. Therefore, in the Heisenberg picture, they will also have mean zero.
The  variance of these operators, are
\begin{equation}\label{eq.valorx2p201}
\begin{split}
 \Delta \thickhat{x}^2_{H} &=  \langle \thickhat{x}^{2}_{H} \rangle =  \frac{E_{0}}{m}\left(\frac{u^{2}}{w^{2}_{0}}+\bar{v}^{2}\right),\\
\Delta \thickhat{p}^2_{H}&=   \langle \thickhat{p}^{2}_{H} \rangle =m E_{0}\left(\frac{\dot{u}^{2}}{w^{2}_{0}}+\dot{\bar{v}}^{2}\right). 
\end{split}
\end{equation}

The mean value of the Hamiltonian, Lagrangian   and   position-momentum correlation operator  of the system are 
\begin{equation}\label{eq.HamiltQ01}
\begin{split}
    \langle \thickhat{H}_{H}\rangle & = \frac{w}{w_{0}}E_{0} Q^{*} ,\\
        \langle \thickhat{\mathcal{L}}_{H}\rangle &=\frac{w}{w_{0}}E_{0}Q^{*}_{1},\\
    \langle \thickhat{\mathcal{C}}o_{H}\rangle &=\frac{w}{w_{0}}E_{0}Q^{*}_{2},
\end{split}
\end{equation}
where
\begin{equation}
  \begin{split}
    Q^{*} & = \frac{1}{2w_{0}w}\left[w^{2}_{0}\left(\dot{\bar{v}}^{2} +w^{2}\bar{v}^{2}\right)+w^{2}u^{2}+\dot{u}^{2}\right],\\
Q^{*}_{1}&=\frac{1}{2w_{0}w}\left[w^{2}_{0}\left(\dot{\bar{v}}^{2}-w^{2}\bar{v}^{2}\right)    +  \dot{u}^{2}-w^{2}u^{2}\right],\\
Q^{*}_{2} &= \frac{1}{2w_{0}w}\left[2w u\dot{u} +2w w^{2}_{0} \bar{v}\dot{\bar{v}} \right].
\end{split}  
\end{equation}
Furthermore, using the condition of Eq. (\ref{eq.condition001}), we find that  
\begin{equation}\label{eq.relationQQ01}
    {Q^{*}}^{2}-\left({Q^{*}}^{2}_{1} +{Q^{*}}^{2}_{2}\right)=1.
\end{equation}
This relationship is useful to analyze the non-classical behavior of the system through the covariance matrix.

The covariance matrix $\bm{V}$ in terms of the operators  $\thickhat{b}$($\thickhat{b}^{\dagger}$) is 
\begin{equation}
    \bm{V}=\left(\begin{array}{c c}
         n_{H} +\frac{1}{2}& \bar{m}_{H} \\ 
         \bar{m}^{*}_{H}& n_{H} +\frac{1}{2}
    \end{array}\right),
\end{equation}
where the elements of the diagonal are related to the number of photons or particles, i.e., $n_{H}=\langle \thickhat{b}^{\dagger}_{H}\thickhat{b}_{H}\rangle-\langle\thickhat{b}^{\dagger}_{H}\rangle\langle\thickhat{b}_{H}\rangle$ and  the off-diagonal elements   are described as $m_{H}=\langle \thickhat{b}^{2}_{H}\rangle -\langle \thickhat{b}_{H}\rangle^{2}$. We found that
\begin{equation}
\begin{split}
n_{H}+\frac{1}{2}&=\frac{\langle \thickhat{H}_{H}\rangle}{\hbar w },\\
\bar{m}_{H}&=-\frac{1}{\hbar w}\left[\langle \thickhat{\mathcal{L}}_{H}\rangle-i\langle \thickhat{\mathcal{C}}o_{H}\rangle\right].
\end{split}
\end{equation}
Then, the covariance matrix is written as
\begin{equation}
    \bm{V}=\frac{1}{\hbar w}\left(\begin{array}{c c}
         \langle \thickhat{H}_{H}\rangle  & -\langle \thickhat{\mathcal{L}}_{H}\rangle +i \langle\thickhat{\mathcal{C}}o_{H}\rangle \\
         -\langle \thickhat{\mathcal{L}}_{H}\rangle -i \langle \thickhat{\mathcal{C}}o_{H}\rangle& \langle \thickhat{H}_{H}\rangle
    \end{array}\right).
\end{equation}

Analyzing the behavior of the covariance matrix over time, we find that $\bm{V}$  is positive.  The classicality criterion of the states of the system is determined by the condition    $n_{H}>\vert m_{H} \vert $ \cite{englert2003tutorial,Alexssandre2022unravelling}. So, we can define the classicality function as $\mathcal{C} =n_{H}-\vert m_{H} \vert $, that is 
\begin{equation}\label{eq.clasicalidade01}
    \mathcal{C}(\bar{n},Q^{*})=\left(\bar{n}+\frac{1}{2}\right) \left[ Q^{*}-\sqrt{{Q^{*}}^{2} -1 } \right]-\frac{1}{2}.
\end{equation}
where the relation of Eq.  (\ref{eq.relationQQ01}) was applied. The classicality function helps us to identify the classical and non-classical behavior of the system.   The states of the  trapped ion are classical if   $\mathcal{C}(\bar{n},Q^{*})>0$, while for $\mathcal{C}(\bar{n},Q^{*})<0$, the states are non-classical. 
\nocite{*}
\bibliography{apssamp}

\end{document}